# Modelling cell reactions to ionizing radiation - from a lesion to a cancer

L. Dobrzyński[1,*], K. W. Fornalski[1,2], J. Reszczyńska[1], M. K. Janiak[3]

[1] National Centre for Nuclear Research (NCBJ), A. Sołtana 7, 05-400 Otwock-Świerk, Poland

[2] Ex-Polon Laboratory, Podleśna 81a, 05-552 Łazy, Poland

[3] Department of Radiobiology and Radiation Protection, Military Institute of Hygiene and Epidemiology (WIHE), Kozielska 4, 01-163 Warszawa, Poland

* corresponding author

*Abstract*

This paper focuses on the analytic modelling of responses of cells in the body to ionizing radiation. The related mechanisms are consecutively taken into account and discussed. A model of the dose- and time-dependent adaptive response is considered, for two exposure categories: acute and protracted. In case of the latter exposure, we demonstrate that the response plateaus are expected under the modelling assumptions made. The expected total number of cancer cells as a function of time turns out to be perfectly described by the Gompertz function. The transition from a collection of cancer cells into a tumour is discussed at length. Special emphasis is put on the fact that characterizing the growth of a tumour (i.e., the increasing mass and volume) the use of differential equations cannot properly capture the key dynamics – formation of the tumour must exhibit properties of the phase transition, including self-organization and even self-organized criticality. As an example, a manageable percolation-type phase transition approach is used to address this problem. Nevertheless, general theory of tumour emergence is difficult to work out mathematically because experimental observations are limited to the relatively large tumours. Hence, determination of the conditions around the critical point is uncertain.

1. *Introduction*

The development of cancer in the body by transition of normal cells to cancerous ones is a complicated multistep process in which many non-linear processes play a significant role. A living cell is a very complex biophysical system. Radiation-induced adaptive response, the bystander effect, and abscopal effects at low radiation doses and dose rates are the key processes which need to be addressed when modelling radiation carcinogenesis.

According to the classical theory, cancer is initiated by a set of mutations in certain genes in a cell. The mutations arise because of inefficient DNA lesion recognition and/or repair. Unrepaired lesions that result from the DNA replication errors and the activity of metabolic free radicals can produce spontaneous mutations occurring at the rate from about $1\cdot10^{-7}$ to $5\cdot10^{-6}$/gene/cell/year (Steen, 2000). This relatively broad range, which is important for further considerations, may be due to variable output and/or activity of repair proteins. Their activity, in turn, can be modified by exogenous stimuli.

Exogenic lesions, caused, e.g., by exposure to intermediate and high doses of ionizing radiation, may lead to the development of cancer through radiation-induced genetic and epigenetic changes. Within this framework, biological effects of ionizing radiation are modelled by a step-by-step introduction of key processes that lead from single changes in the DNA to a full-blown cancer. In contrast to exposures at intermediate and high radiation doses, absorption of low radiation doses, especially when delivered at low dose rates is unlikely to produce multiple irreparable DNA lesions, but still alerts the DNA damage surveillance system. This results in a stimulated repair of numerous DNA lesions in genes, including those associated with cell replication and metabolism.

In this case, the increased repair capacity of the cells (and therefore of the whole organism in the case of whole body exposure), which is manifested by a decreased overall rate of fixed mutations in the DNA, can translate into reduced risk of neoplastic transformation of cells and of cancer development. The degree of natural protection stimulated by low radiation doses depends on the type of radiation, its dose, and dose rate.

Within a standard adaptive response study design, a small *priming dose* is used to up-regulate adaptive-response mechanisms (which represent a mild stress response and *in vivo* can involve a hierarchy of natural protective mechanisms (Scott, 2014, 2017). A large *challenging dose* is then administered (usually shortly after the priming dose) and biological endpoints (e.g., the rates of cell deaths, mutation or neoplastic transformation) of such a combined exposure are compared with the ones when only the challenging dose is used. A reduced frequency of adverse biological effects in the presence of the priming dose indicates a rapid adaptive response (i.e., rapid adaptation) induced between the two exposures; which may involve epigenetic changes (Scott et al., 2009). The priming dose can be brief or protracted for these effects to take place. In addition, the priming dose can lead *in vivo* to reduction in the rates of mutations (Ogura et al., 2009) and neoplastic transformation (Redpath et al., 2001), to a level below the spontaneous frequency; presumably as a result of up-regulation of the body's natural defenses (Scott 2014, 2017).

An important role in cancer formation may be played by close-by and distant cells in tissue through intercellular signalling. These signals are responsible for induction of bystander (nearby cells) and abscopal (distant cells) effects. However, detailed mechanisms of these effects have not yet been fully resolved and there are inconsistencies in their understanding. For example, some authors reported that in studies of the bystander effect, signals from irradiated cells to a unirradiated cells surrounding an irradiated one exacerbate lesions in the latter (Marin et al., 2015; Marcu et al., 2009). However, other studies demonstrate elimination of such lesions (e.g. Mothersill and Seymour 2005) through the induction of apoptosis (e.g. self-destruction of both hit and not hit transformed cells). The latter is now considered a different form of adaptive response and relates to stimulated up-regulation of natural protective mechanisms (Scott 2014). In the case of the bystander effect one can try to model it (e.g. Khvostunov and Nikjoo 2002; Hattori et al. 2015) and use Monte Carlo simulations (e.g., as in the paper by Fornalski et al. (2017), Hattori et al. (2015)), and the result will clearly depend on the employed model. It is believed that mechanisms of bystander effects and adaptive response (commonly associated with low dose exposures) are basic components of the cellular homeostatic response (Marcu et al., 2009). Adaptive response to radiation has been described theoretically and modelled in a number of ways. However, none of the currently available quantitative risk models have covered the path from the deposition of radiation energy in a cell to a developed (i.e. neoplastically transformed) cancer cell. As will be discussed in section 6, in order to fully describe this process, it is evident that any change from a simple collection (set) of cancerous cells to a more complex system of a malignant tumour must be associated with a basic reorganization of the initial set into a new entity, the properties of which cannot be readily and uniquely derived from the properties of the initial system. Until a more precise understanding of the underlying mechanisms is reached, we will focus only on description of hit cells. Thus, one has to accept the fact that the calculations presented in this paper will have to be corrected for the two aforementioned effects once their mathematical form and relative strengths are worked out.

Describing (or modelling) the development of radiogenic cancer from the initial radiation-induced/exacerbated genetic and/or radiation-exacerbated genetic and/or epigenetic changes to a clinically detectable neoplasm may be regarded as an unattainable task. Moreover, since there are different ways for cancer to arise, different conceptual models have to be introduced. One of these models was introduced by Hanahan and Weinberg in their two seminal papers (2000; 2011). Their conceptual model focused on the hallmarks of cancer which provide guidance on the key processes that should be addressed when developing a quantitative, mechanism-based model of the development (risk) of cancer.

We propose a possible biophysical interpretation of the processes of creation of radiation-induced changes in the DNA and the ensuing mutations in exposed cells. Transformation of a mutated cell into a neoplastic one is also discussed. In all these processes, the adaptive response mechanism (which has been proposed and successfully used in earlier theoretical studies) is implemented and discussed. Most of the mathematical formulae will be presented as proper probability functions which can be used in Monte Carlo simulations.

The paper is organized as follows: in section 2, we address the possible outcomes of exposing a cell to ionizing radiation. In section 3 the path from lesions to mutations is described, mainly basing on the Random Coincidence Model - Radiation Adapted (RCM-RA) (Fleck et al. 1999). Section 4 describes adaptive responses of cells exposed to acute or protracted (continuous) irradiation. This leads to considerations of transformation of a mutated cell into a cancer cell, including the relationships between the dose-rate and the number of mutations on the number of the developed cancer cells (Section 5). A more detailed description of the inception of a tumour from pre-cancerous cells is the subject of Section 6. Conclusions are presented in the final section of the paper.

The general idea of this modelling utilized by us is presented on the flow-chart below, Fig.1. Every step is described in each section of the paper.

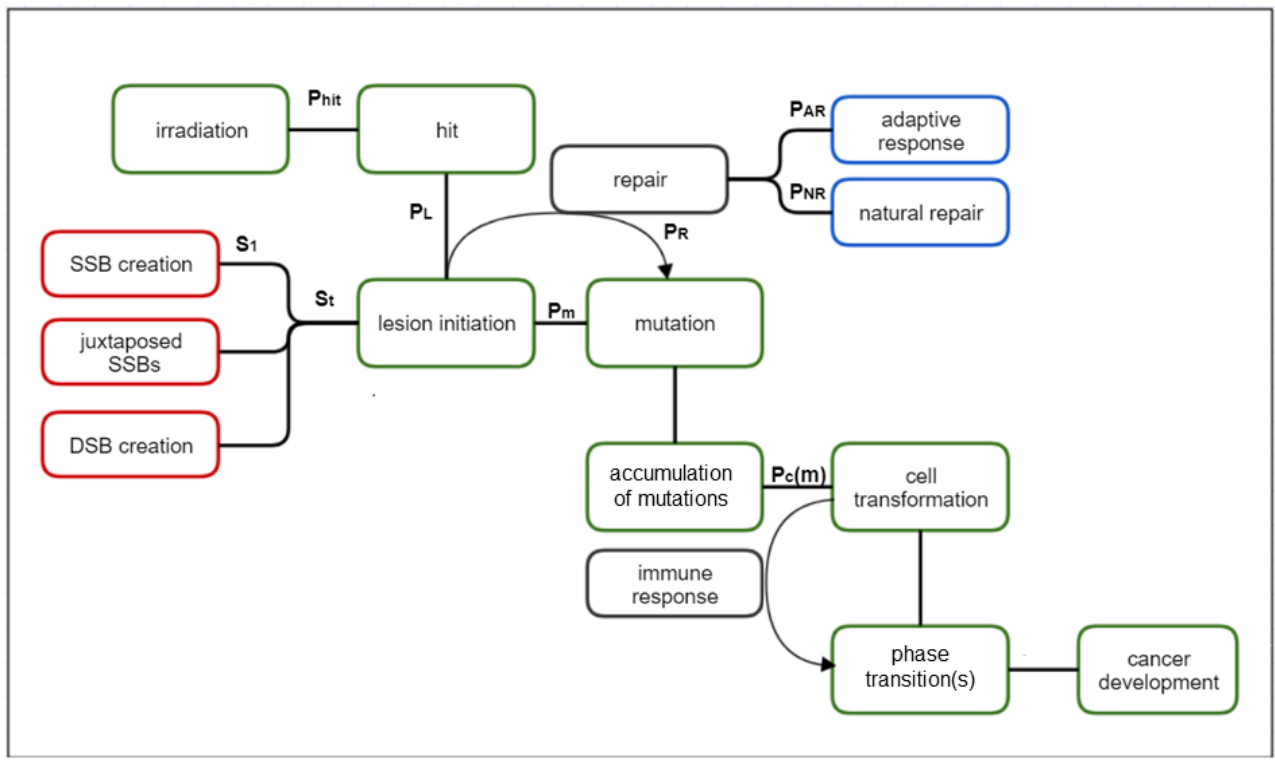

Fig.1. Flow-chart of the model used. The meanings of symbols are explained in the Table 1 and in the text of the paper.

Table.1. Summary of main probability functions in the presented model.

| Probability function | Described process or explanation |
|---|---|
| $P_{hit}$ | radiation DNA hit in a cell |
| $P_L$ | Creation of the DNA damage from the hit |
| $S_1$ | initiation of single base change or a SSB per unit of time |
| $S_t$ | any of all considered types of DNA breaks occurring in time |
| $P_m$ | mutation creation per unit of time |
| $P_R$ | repair process, reduction of number of lesions per unit of time |
| $P_{AR}$ | radiation-activated mechanism of the adaptive response |
| $P_{NR}$ | natural repair of DNA lesions |
| $P_c(m)$ | cell transformation into a cancer cell due to number of accumulated mutations in time |

2. *Creation of lesions in a cell after deposition of radiation energy*

The interaction of ionizing radiation with matter depends on the type *R* of radiation, and on its energy, *E.* This interaction is typically described by the cross section for a given process, $\sigma(R,E)$. The cross section of interaction of ionizing radiation with matter (or a cell) depends on many physical effects. As an example, an interaction of a single photon with matter is briefly described in Appendix A.

Ionization events in the cell may lead to several biophysical effects. Here, only lesions that may lead to neoplastic transformation will be considered, such as radiation-induced single (SSBs) and double strand (DSBs) breaks will be considered.

As an example, for electrons of energies 5.6 eV and 9.6 eV, the values of SSB cross-sections are $2.4 \cdot 10^{-14}$ cm$^2$ and $4.7 \cdot 10^{-14}$ cm$^2$, and those of the DSB cross-sections are $4.1 \cdot 10^{-15}$ cm$^2$ and $4.5 \cdot 10^{-15}$ cm$^2$, i.e. the former are by an order of magnitude smaller than the latter (Chen et al. 2016).

In addition to the cross sections, one needs also to consider the probability of a radiation hit at the DNA, $P_{hit}$, which depends on the flux of impinging particles, and the surface density (number per cm$^2$) of the elementary objects (e.g. the DNA or its secondary structures) to be hit. The probability of a hit can be combined with the dose $D$ (i.e., dose rate, $\dot{D}$, multiplied by time of exposure) absorbed by the object. Such a dependence may be linear at a low dose; however one could postulate that it must saturate at a high dose. Thus, the total probability function describing the creation of a DNA damage (lesion) after the radiation hit, $P_L$, can be described as:

$$P_L = A\sigma P_{hit} = A\sigma \left(1 - e^{-cD}\right), \tag{2.1}$$

where $A$ is a normalization constant, and $c$ denotes a scaling constant. Because cross sections are expressed in barns b (1b = $10^{-24}$ cm$^2$), constant $A$ must be expressed in cm$^{-2}$. Obviously, it is the product of the thickness of the target (in cm) and the numerical density of the interacting objects, (e.g. number of cells per cm$^3$). The concept of $P_{hit}$ in eq. (2.1) was originally used by us in the Monte Carlo chain cellular model (Fornalski et al. 2011), but the validity of this formula has not been verified. Now, such a validation is presented in Appendix B.

3. *Phenomenological descriptions of lesions and mutations in irradiated cells*

When a damage to the DNA is identified as a single lesion with probability $P_L$, one can consider the probability of creating a mutation resulting from an unrepaired or improperly repaired lesions. The process of mutation creation may be described by a polynomial, presumably dependent on the linear energy transfer (LET) of a given type of radiation $R$ (Kellerer and Rossi 1976). In particular, one can assume that the probability of a mutation caused by a mixed radiation field may have the form:

$$p_1 = 1 - e^{-\sum_R \sum_{i=0} a_{i,R} D_R^i}, \qquad (3.1)$$

where $D_R$ represents an absorbed dose of radiation $R$, and $a_{i,R}$ are experimentally derived parameters for a given type of radiation $R$. Obviously the units of $a_{i,R}$ must ensure the dimensionless product in the exponential function. At low doses eq. (3.1) can be approximated by:

$$p_1 = \sum_R \sum_{i=0} a_{i,R} D_R^i. \qquad (3.2)$$

Form (3.2) is used by International Atomic Energy Agency in biodosimetric standards (IAEA 2011) where aberrations (such as dicentrics) are used to assess the dose received by the irradiated person. According to conclusions by Kellerer and Rossi (1976) and Szłuińska et al. (2005), this probability ($p_1$) is linear for high LET radiation, such as neutrons ($i = 1$), and linear-quadratic ($i = 1, 2$) for low LET radiation, such as gamma- or X-rays.

A different approach was proposed by Fleck et al. (1999) who presented a biophysical model initially considering the probabilities of SSBs and DSBs induced in the DNA either one after another (SSBs) or during a massive attack of reactive oxygen species (ROS) or ionizing radiation (DSBs). In their RCM-RA model, the dose-rate serves as a crucial parameter in the role of ionizing radiation in this process. Assuming that the metabolic chemical burden production rate is $C$ (per unit time) and the dose rate is $\dot{D}$, the probability of a single base change or a SSB per unit of time is (original abbreviations):

$$S_1 = \alpha C + \beta \dot{D}, \qquad (3.3)$$

where $\alpha$ and $\beta$ are weighting factors. The first term on the right side of eq. (3.3) describes the probability of damage (per unit time and a nucleotide) arising from the natural metabolism. The second term describes a similar effect, caused by ionizing radiation. The $\beta$ constant (in $Sv^{-1}$ times the time unit) has been also calculated (for low LET radiation only) by Fleck et al., 1999; see their eq. (A.10). The $\alpha$ coefficient is dimensionless.

Let the average time needed for error-free repair be $\tau$. It may be expected that this must depend on the efficiency of repair enzymes. The average number of repaired lesions within $\tau$ is:

$$S_2 = (\alpha C + \beta \dot{D})\tau. \qquad (3.4)$$

Consequently, the probability per unit time of the development of a DSB as a result of the sequential production of two SSBs closely related in space and time, should be proportional to:

$$S_t = (\alpha C + \beta \dot{D})^2 \tau. \qquad (3.5)$$

In this model (Fleck et al., 1999) the rate of the poorly repairable DSBs is linear in the time needed to repair single lesions. Even if this assumption may not necessarily hold true, it may be accepted as a first approximation. The kinetics of a DSB repair in humans was recently considered in greater detail by Jain et al. (2017) who demonstrated that irradiation at low dose rates increases the efficiency of such a repair. This finding is important in view of the level of complexity of such repair of the naturally created DSBs. This involves a synchronized action of dozens of proteins involved in the two repair pathways, homological and non-homological, occurring with different accumulation speeds. In the meta-analysis carried out by Kochan et al. (2017), the kinetics of the DSB proteins behaves with time $t$ as $1 - \exp(-t/\tau_1)$, where the characteristic time $\tau_1$ is the inverse accumulation speed, which could be modified by adaptive mechanisms.

Assuming that the production of the repair enzymes increases up to the saturation at a certain equilibrium, Fleck at al. (1999) argue that it is reasonable to assume that the average repair time must decrease in a manner inversely proportional to $(1 + \delta \dot{D})$, where $\delta$ is a coefficient related to the enzyme production rate, possibly dependent on LET. This assumption stems from the following reasoning: the higher the radiation dose rate, the greater the number of the induced repair enzymes per time unit. With the elevated efficiency of the repair enzymes the probability of their presence in a close proximity to a damaged DNA fragment must increase. Hence, the average repair time of the damaged fragments should decrease with increasing dose rate. Eq. (3.5) will then take the form:

$$S_t = \left(\alpha C + \beta \dot{D}\right)^2 \frac{\tau_0}{1+ \delta \dot{D}}, \qquad (3.6)$$

where $\tau_0$ denotes the characteristic repair time for the non-radiation-induced lesions. At a sufficiently large $\alpha C$ this equation exhibits an apparent hormetic-like dip of $S_t$ at the low dose rate exposure:

$$\dot{D}_{min} = \frac{\delta\alpha C - 2\beta}{\beta\delta}. \tag{3.7}$$

Instead of reliance on the postulated eq. (3.6), one could alternatively assume:

$$S_t = (\alpha C + \beta\dot{D})^2 \tau_0 e^{-\delta\dot{D}}, \tag{3.8}$$

which, at small values of $\delta\dot{D}$, is not very different from (3.6). Eq. (3.8) represents a reverse situation: instead of observing the minimum, the function (3.8) exhibits a maximum at the value:

$$\dot{D}_{max} = \frac{2\beta - \delta\alpha C}{\beta\delta}. \tag{3.9}$$

This shows that the final description of $S_t$ is very sensitive to assumptions, so one should be careful with postulating a definite formula for the dependence of repair time on the dose-rate. In fact, the main assumption which led Fleck et al. (1999) to propose such a dependence (3.6) was that production of the repair enzymes increases linearly with dose rate, which may not necessarily be the case. To conclude, we note that eq. (3.7) offers a limit on the coefficient $\delta$, namely $\delta$ must be larger than or equal to $2\beta/(\alpha C)$ if one accepts the shortening of $\tau$ with the dose rate as in eq. (3.6), and less than $2\beta/(\alpha C)$ if one accepts eq. (3.8). Apparently, since the number of DNA lesions should initially increase and decrease only after maximal accumulation of the repair enzymes (Kochan et al, 2017) eq. (3.8) could also be accepted based on such phenomenological considerations.

Let us note that at very low dose rates the $S_t$ values behaves as:

$$S_t \sim const\,(1 - \delta\dot{D}), \tag{3.10}$$

where $const=(\alpha C)^2 \tau_0$. Let us also note that the positive value of $(\dot{D})_{min}$ may equally well bind any of the four constants appearing in the eq. (3.7). The hormetic-like minimum observed in Fig.1 of Fleck et al. (1999) can be explained based on the assumption of reduction of repair time with the dose-rate as in eq. (3.6) without considering that the repair time may vary with time after irradiation. Last but not least, at high dose rates, $S_t$ becomes either proportional to the dose rate, if eq. (3.6) is used, or tends to zero, if eq. (3.8) is used. In the former case, one observes a linear no-threshold (LNT) behaviour while eq. (3.8) demonstrates the decreasing probability of DSBs production due to the shortening of the time of the first SSB repair. In such a situation, a DSB would be expected to mainly be produced by a mechanism different than the consecutive induction of two juxtaposed SSBs. At very low dose rates or at small value of $\delta$ the use of either of equations, (3.6) or (3.8), practically yields the same results.

Fleck et al. (1999) assumed that the repair of double lesions of all possible kinds cannot be successful (is error-prone) and that the rate of their appearance is proportional to the dose rate. With this assumption, Fleck et al. (1999) showed that the proportionality constant has the form $(1/f_{nuc})\overline{z_F}\beta^2$, where $f_{nuc}$ denotes the average fraction of the volume of a cell nuclei (about 0.3), and $\overline{z_F}$ is the mean specific energy per event deposited in a critical volume. Thus, the modified $S_t$ of eq. (3.8) is:

$$S_t = \left(\alpha C + \beta \dot{D}\right)^2 \tau_0 e^{-\delta \dot{D}} + (1/f_{nuc})\overline{z_F}\beta^2 \dot{D} \ . \qquad (3.11)$$

It is easy to check that the second term on right-hand side of this equation has the dimension of the inverse of time, e.g., 1/s. Eq. (3.11) can still be supplemented by a term accounting for cell killing. Again, this model leads to LNT-type relationship at high dose rates and should not be used in the considered range of dose rates. In the modification of the RCM-RA model (Schöllnberger et al., 2001) published two years later after the report of Fleck et al. (1999) the average deposited energy $\overline{z_F}$ was substituted by $\overline{z_F}^+ = \frac{D}{1-e^{-D/\overline{z_F}}}$, where *D* denotes the person's lifetime dose (whether it includes natural background radiation is not clearly stated in the cited paper), and $\overline{z_F}^+$ corresponds to the 'mean of the specific energy deposited in the affected cell volumes, *i.e.*, volumes that have experienced at least one energy deposition event'.

Eq. (3.11) can be modified by taking SSBs into account, using the same notion as in eq. (3.2) (or eq. (3.1) in general). However, SSBs are usually efficiently repaired, and the unrepaired DSBs dominate, so, for practical reasons, there is no need to make such a modification.

Using eq. (3.11) with the first term as in eq. (3.6) ~~only~~, the authors of the RCM-RA model (Fleck et al. 1999; Schöllnberger et al.2001) obtained an almost perfect fit to the Cohen's data (Cohen 1995), corrected for smoking, which means that within the scope of the cited papers SSBs appear to play a minor role in the creation of mutations and the ensuing neoplastic transformations of cells. The authors of the RCM-RA model got a relatively shallow minimum of the lung cancer mortality after exposures at a very low dose rate (roughly 1 mGy/y) and the apparently steady increase at higher dose rates. Such an increase has been attributed to the differential inhibition of the body's natural anticancer defences (Scott et al, 2009). Within the indicated paper (which introduces a hormetic relative risk [HRR] model), stochastic thresholds for inhibition apply. In view of some other reports on the effects of radon exposures (Dobrzyński et al. 2018), this increase in the considered range of doses/dose rates is questionable.

For a more complete description of the development of a mutation one needs to include the probability of the creation of a lesion, $P_L$, which is represented by eq. (2.1). As mentioned earlier, some additional appropriate repair mechanisms need to be accounted for in that modified formula as well, to allow for a reduced probability of the creation of a mutation. Only after considering the expected reduction in the number of lesions by the repair of the DNA damage sites (such as SSBs and DSBs) one can reliably calculate the expected number of point mutations. In addition, as discussed below, one mutation is not likely to produce a cancer cell. Thus, the joint probability function of mutation's creation per unit of time can be presented as:

$$P_{mutation} = P_L \cdot p_1 \cdot (S_t - P_R), \tag{3.12}$$

where $p_1$ represents probability of stable mutation, $P_R$ is the general probability function (per unit of time) that describes repair mechanisms additional to the one already used in $S_t$ (time dependent probability of creation of unrepairable DSB lesion with included repair mechanism of SSB). Let us recall that probabilities $P_L$ and $p_1$ are dimensionless. Eq. (3.12) is consistent with (but different from) the Feinendegen's model of dual action (Feinendegen et al. 2010, 2012). The probability of the occurrence of detrimental effects, $S_t$, can be described by eq. (3.11) or by other forms, like the one presented by Dobrzyński et al. (2016).

The repair probability, $P_R$, can be generally composed of two main components ($P_R = P_{NR} + P_{AR}$): natural repair of the DNA lesions (cell age dependent and possibly genetically determined), $P_{NR}$, and the radiation-activated mechanism of adaptive response, $P_{AR}$. That these two repair mechanisms can be simply added is the assumption only, valid when the radiation-induced repair is not making use of the natural protective mechanisms. If this is not the case, $P_{AR}$ should be considered as $P_{NR}(1+R_I)$, where $R_I$ is describing stimulation of the natural protection due to radiation.

$P_{NR}$ can likely be approximated by the inverted sigmoidal function (the Mehl-Avrami equation), where the repair possibilities decrease over time:

$$P_{NR} = C\, e^{-a\, K^n}, \tag{3.13}$$

where *C* and *a* are scaling constants, and *K* is the normalized age of the irradiated cell, i.e. the real time divided by an accepted characteristic time constant. The parameter *n* is of crucial importance because it is determined by down-regulation of the repair enzymes with the cell's age.

A large fraction of unrepaired cells also undergo mitotic death or apoptosis (programmed cell death), and thus do not contribute to the mutation load (Bauer 2007). In the context of oncogenesis, both types of cell death offer a 'successful' resolution of ineffective repair of the DNA damage, especially after a small number of dead cells can be tolerated. Moreover, it seems that removal of already transformed cells is also possible after irradiation at low doses through intercellular apoptotic signalling (Bauer 2007).

4. *Adaptive response to ionizing radiation*

The adaptive response phenomena include triggering of repair mechanisms, especially in the DNA, after irradiation of cells, tissues, or a whole organisms. The number and efficiency of the activated repair enzymes are associated with the number of ionization events, which depend on the dose and dose rate. Adaptive response is assumed to be reasonably well accounted for using equations similar to eq. (3.1) or (3.2). The main implication of the latter is that the efficiency of repair is expected to grow continuously with the dose of radiation. However, since the effectiveness of repair saturates at high doses, this equation is expected to be reliable only for low doses.

Additionally, one can assume that the efficiency of the repair enzymes decreases exponentially with increasing dose. In what follows we shall use an exponential decay in which:

$$p_2 = e^{-bD}. \tag{4.1}$$

Finally, the dose/dose-rate related probability of the effectiveness of the repair enzymes can be described as a product of eq. (3.2) and (4.1):

$$P(D) = p_1 p_2 = \left(\sum_R \sum_{i=0} a_{i,R} D_R^{\,i}\right) e^{-bD}. \tag{4.2}$$

As already mentioned, all equations (3.1), (3.2), (4.1) and (4.2) are connected with the dose/dose-rate relation.

With respect to the time dependence one can assume, as the first approach, that the number of repair enzymes and their effectiveness increases with time (after an initiating event) with a probability:

$$p_3 = \mu_0 + \mu\, t, \tag{4.3}$$

where $\mu$ describes the enzymes production rate after a pulse of radiation. This assumption should be not far from reality especially at short times after the irradiation.

If one considers a single radiation pulse only, the effectiveness of the activated enzymes, after the initial rise in their concentration, must also decrease with time with a certain time constant (lifetime), $1/\lambda$ (IAEA 2011). Were the probability of such a decrease per unit of time constant, this decrease would be described by:

$$p_4 = e^{-\lambda t} \;, \tag{4.4}$$

which finally would lead to the overall time dependence:

$$P(t) = p_3 p_4 = (\mu_0 + \mu t)e^{-\lambda t} \;. \tag{4.5}$$

The general shape of eq. (4.5) is similar to the shape of eq. (4.2) and can be generalized in an analogous manner:

$$P(t) = \left(\sum_{n=0} \mu_n t^n\right)e^{-\lambda t} \;, \tag{4.6}$$

where the index *n* may be of non-integer type as the proportionality of $p_3$ with time still remains an arbitrary assumption. For practical reasons, however, the simplified form given by eq. (4.5) is preferred.

Finally, the joint probability function of the adaptive response should be dependent both on the dose-rate, $\dot{D}$, and the time, *t*. Obviously the product of these two parameters is the absorbed dose. One should also note that at high doses the consideration of a time-dependent adaptive response makes no sense because of the smallness or non-existence of adaptedness (Fornalski 2014). In numerical calculations one introduces time steps, $k \in \{1,\ldots, k_{max}\}$ and the dose per unit time step (*D*), i.e. the dose rate rather than the dose. The value of the time step has to be chosen independently. It seems convenient to use the time step equal to $\tau$ as introduced in eq. (3.4). As indicated above, both variables can be used independently in two different equations, depending on the context. Thus, the simplest forms of the appropriate functions are:

$$P(D) = \alpha_1 D^n e^{-\alpha_2 D} \;, \tag{4.7}$$

$$P(k) = \alpha_4 k^m e^{-\alpha_3 k} \;. \tag{4.8}$$

Let us note that the normalization constants, $\alpha_1$ and $\alpha_4$, are dependent on the remaining parameters $n$ and $m$ (higher than 1 to obtain a hunchbacked shape of the curves), so that $\alpha_1 = \alpha_1(n, \alpha_2)$, and $\alpha_4 = \alpha_4(m, \alpha_3)$. The true dependence is determined by the assumed ranges of $D$ and $k$, respectively. This approach was successfully used in the Monte Carlo modelling where the joint probability function of the adaptive response was calculated in a discrete form (Fornalski 2014; Dobrzyński et al. 2016; Fornalski et al. 2017):

$$P_{AR} = C \sum_{k=0}^{K} D^n (K-k)^m e^{-\alpha_2 D - \alpha_3 (K-k)} \; , \tag{4.9}$$

where $C$ represents a normalization constant and $K$ – the cell's age given as the number of elementary time steps. This equation may be written in a continuous form (Fornalski 2014; Dobrzyński et al. 2016; Fornalski et al. 2017):

$$P_{AR} = C \int_{t=0}^{T} \dot{D}^n (T-t)^m e^{-\alpha_2 \dot{D} - \alpha_3 (T-t)} \, dt \; . \tag{4.10}$$

Let us note that $D$ in eq. (4.9) denotes the dose per time step, whereas the dose rate in the continuous form (4.10) means the dose per unit of time. Obviously such a modification requires the appropriate change of interpretation of the coefficients $\alpha_2$ and $\alpha_3$.

The subtle point in calculations is that one should distinguish whether the dose was delivered in a single step or continuously over a period of time. If the dose is delivered in the $l$'th step only, its effect at the $h$'th time step will be described by the simplified version of eq. (4.9). However, if the dose is delivered continuously from the time $h_0$ to time $h_1$, the situation at each time step becomes more complicated and for the $v$'th time step:

$$P(D,\text{v}; h_0, h_1) = C\, D^n e^{-\alpha_2 D} \sum_{h_0}^{h_1}(\text{v}-h)^m e^{-\alpha_3(\text{v}-h)} \quad , \tag{4.11}$$

where summation runs over $h$ and $\text{v} \geq h_1$ and the dose $D$ should be understood as a constant dose/step, i.e., effectively, the dose rate. If the time step is small enough, the sum of the discrete values on the right-hand site of eq. (4.11) can be changed to an integral, as in eq. (4.10):

$$P(D,t) = C\, D^n e^{-\alpha_2 D}\, I(t) \quad , \tag{4.12}$$

where $D$ denotes a single dose pulse delivered time $t$ ago. The appropriate formulas of the function $I(t)$ are given in Appendix C.

In experiments like those carried out by Jain et al. (2017) the time of observation after the irradiation was close to $\tau$ or not more than a few times longer.

The above considerations are important if one wants to characterize specific situations in regions with the elevated background radiation. In the aforementioned paper by Jain et al. (2017) the level of background radiation was regarded as a priming dose relative to the additional challenge dose to the cells. The first dose was absorbed during chronic (environmental) irradiation, whereas the second dose (up to 2 Gy) was applied over 0.5-2 minutes, i.e., in a much shorter time than the one needed for the development of any adverse reactions as well as of repair mechanisms. In a typical experiment (Shadley and Wolff 1987; Shadley et al., 1987) demonstrating the adaptive response in cells both priming and challenging doses were acute, i.e., applied within a short period of time.

Fig. 2 shows a typical priming-dose effect as a special example of the adaptive response (as modelled by us, eqs 4.7-4.8), when *m*=1 for two irradiation times. For ease of the comparison, both curves were normalized to the same maximum. Fig. 2 displays the case of *m*=2 (eqs 4.7 and 4.8) and the irradiation applied between the 2nd and 20th time step. One can note the qualitative behaviour of this response vs. time which is not much different from the assumed response in each time step. During the irradiation the response smoothly increases with every time step, but does not saturate, indicating that the assumed model may not work. If it worked with chronic irradiation (e.g. during environmental exposures), we should grow more resistant to it with age (the probability of adaptation saturates at older age). Obviously, our own immunological fitness deteriorates with time, so this effect must be included in such considerations. The problem is resolved when the calculated irradiation time increases. Fig. 3 shows response to the irradiation time 5 times longer than in the case shown in Fig. 2. The response is apparently flattening out and decreases relatively soon after discontinuation of the irradiation. Such a dependence shows that chronic exposures cause a constant adaptation of the organism to radiation which was also demonstrated in the earlier Monte Carlo studies (Fornalski 2014). Although the strength of the maximal adaptive response is limited, it can still be substantial. Therefore, as observed by Jain et al. (2017), inhabitants of regions with a substantially elevated background radiation can indeed present with a higher radio-resistance.

If the dose-rate becomes too high for the enzymes to perform the error-free repair, the constant parameter related to the dose-rate (e.g. *b* in eq. (4.1)) should have the meaning of the inverse of characteristic dose-rate which describes the effectiveness of the enzymes. In a more restrictive reasoning one should bear in mind that the formula like (3.1) may be different for low and high dose-rates. We know that different groups of genes are involved in repair actions in these two regimes, so to stay on the safe side one has to limit our considerations to low dose-rates. Thus, the fundamental background of the adaptive response effect is described by eqs (4.7) and (4.8), with the most general form of:

$$P(\xi) = a\, \xi^n e^{-\lambda \xi}, \qquad (4.13)$$

where $\xi$ may denote the dose, dose-rate, as well as the time. Were the validity of eqs (3.1) and (3.2) questioned, eq. (4.13) would still look reasonable. The hunchbacked shape of eq. (4.13), is commonly encountered in the literature. For example, Feinendegen found that the probability of the induction of adaptive response should be given by the probability distribution function with the maximum at low doses and the strongest effect being apparent after some period of time (Feinendegen 2016). The shape of this simple function is governed by two parameters, *n* and $\lambda$, only.

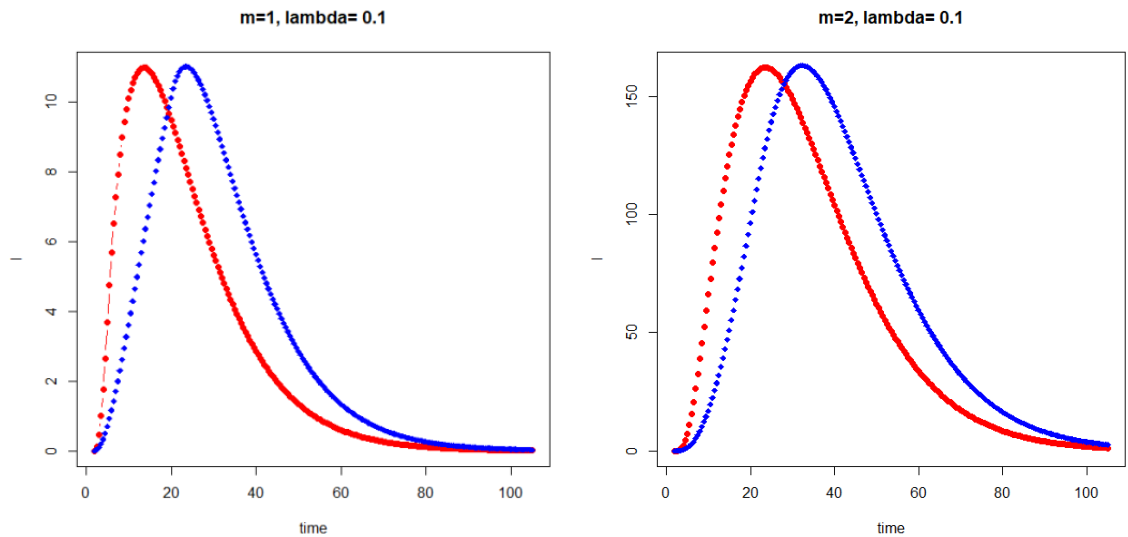

Fig.2 Normalized adaptive response (as modelled by us, eqs 4.7-4.8). Irradiation time 2-5 steps (red), and 2-20 steps (blue) for m=1 (figure on the left) and m=2 (figure on the right).

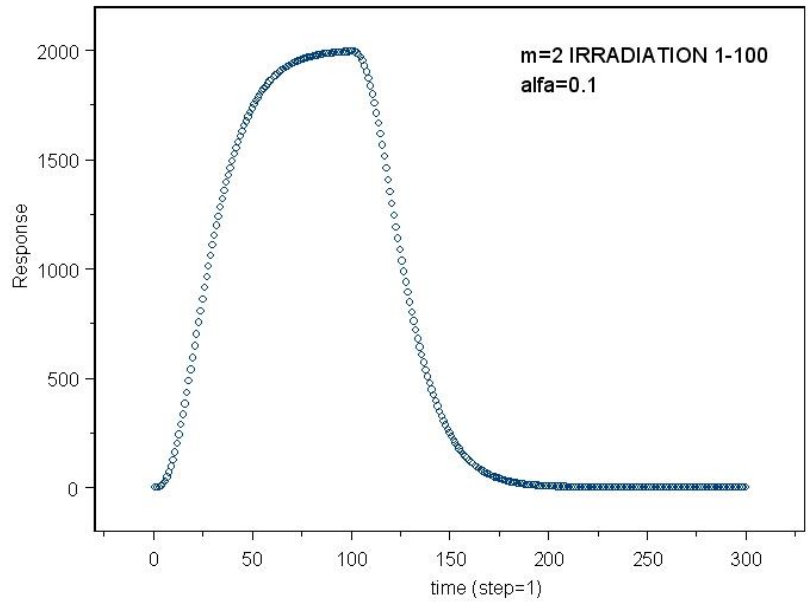

Fig.3 Same as Fig.2 for longer irradiation time (1-100 time steps).

The maximum value of eq. (4.18) is attained at:

$$\xi_{max} = \frac{n}{\lambda},  \qquad (4.14)$$

where it reads:

$$P_{max} = a \left(\frac{n}{\lambda e}\right)^n. \qquad (4.15)$$

In a special case of chronic irradiation, one can easily calculate the mean lifetime or the mean survival fraction of the repair enzymes (or their effectiveness):

$$<\xi> = \frac{\int_0^\infty \xi^{n+1} e^{-\lambda \xi} d\xi}{\int_0^\infty \xi^n e^{-\lambda \xi} d\xi}. \qquad (4.16)$$

Because:

$$P_{total} = \int_0^\infty a\xi^n e^{-\lambda \xi} d\xi = a \frac{n!}{\lambda^{n+1}}, \quad (4.17)$$

the mean lifetime $<\xi>$ becomes equal to:

$$<\xi> = \frac{n+1}{\lambda} \quad (4.18)$$

One can note that the chronic low-rate irradiation can be treated as an infinite series of small radiation pulses. Indeed, integrating the sequence of (4.13) from time zero to infinity after the irradiation time, one obtains:

$$P_t = \alpha \int_0^\infty (t+\theta) e^{-\lambda(t+\theta)} d\theta = \alpha \frac{1+\lambda t}{\lambda^2} e^{-\lambda t}. \quad (4.19)$$

This result shows that in spite of a linear increase of total dose at longer times this probability exponentially decreases with time with the rate of the initial reaction to dose, i.e. $1/\lambda$. Initially, the probability increases, although non-linearly. The same effect is observed when the leading coefficient in eq. (4.13) is changed to $t^2$. Then $P_t$ changes to:

$$P_t = \alpha \int_0^\infty (t+\theta)^2 e^{-\lambda(t+\theta)} d\theta = \alpha \frac{2+2\lambda t+\lambda^2 t^2}{\lambda^3} e^{-\lambda t}. \quad (4.20)$$

5. *Neoplastic transformation of mutated cells*

While the description presented above was given in terms of dose rates, the cumulative dose itself can be considered as well. In fact, so far the only need for the time variable so far has been to address repair of individual SSBs and the adaptive response. This repair time is stochastic rather than deterministic and is relatively short (according to Fleck et al. (1999), it takes about 40 minutes), so one speak about very low doses when one considers low dose-rates. At such doses the epigenetic term in, e.g., eq.

(3.3) dominates and therefore the second term on the r.h.s. of eq. (3.10) can make a difference. This term strictly relates to a specific cell response to irradiation: production of the hard to repair DSBs. This response, however, even if happens in individual cells, in tissues should also strongly depend on the time elapsed since the irradiation. At a constant dose rate, the number of the repair-resistant DSBs should increase with time, as should the number of the mutated cells. In eq. (3.10) the second term reflects the LNT approach. Thus, one must take into account that the organism counteracts a defective DSBs and other lesions in tissues using repair mechanisms (natural and adaptive responses), as proposed in eq. (3.12).

Neglecting cooperation between cells, Fleck et al. (1999) suggested that the time dependent generation of cells with the 1st mutation (the number of cells per person at time *t* which incurred 1st mutation, $M_1$) should be governed by the equation:

$$\frac{dM_1}{dt} = (B_0 M_0 - B_1 M_1) P_m \tag{5.1}$$

where $P_m$ denotes $P_{mutation}$, see eq. (3.12). $M_0$ in this equation denotes the number of non-mutated cells, while $B_0$ is interpreted as the 'number of critical DNA bases in critical codons of all tumour associated genes per cell'. According to the Human Genome Project[1], a human genome contains about 25,000 coding genes composed of approximately 3 billion DNA base pairs. It seems that the genome includes 291 cancer-associated genes and more than 1% of all genes are thought to be involved in carcinogenesis (Futeral et al. 2004). Hence, about 1% of all the DNA bases are likely to represent such a critical value of $B_0$.

---

[1] www.genome.gov

For cells of the same tissue one can assume that $B_1$ (the value similar to $B_0$, but after the first mutation) should not be much different from $B_0$. It would seem that there should be a minor error if both of these coefficients were substituted by a single one, $B = B_0 = B_1$. Let us note that in the original formulation by Fleck et al. (1999) the last multiplier on the r.h.s of eq. (5.1) is $S_t$. In order to preserve our reasoning, this function was replaced by the probability of mutation, $P_m$, which is much closer to reality.

The solution of differential equation (5.1) is:

$$M_1 = M_0(1 - e^{-BP_m t}) , \qquad (5.2)$$

which shows at small values of time a linear growth of $M_1$ with time (as in eq. (B37) in (Fleck et al, 1999)) and at high *t*-values a saturation (equilibrium), hence $M_1 = M_0$. The saturation, however, most likely overestimates the number of single mutated cells as $M_1 = M_0$ means that the number of mutated cells is equal to the number of all cells.

To find the expected number of cells with two mutations Fleck et al. (1999) consistently suggests equation similar to eq. (5.2):

$$\frac{dM_2}{dt} = (M_1 - M_2)P_m . \qquad (5.3)$$

The solution is:

$$M_2 = M_1(1 - e^{-BP_m t}) = M_0(1 - e^{-BP_m t})^2 . \qquad (5.4)$$

With the increase of time the number of such cells must be smaller than $M_1$. Equation (5.4) can next be easily generalized to the case of *m* mutations per cell (see eq. (5.6)). It is important to note that according to this procedure the number of mutated cells grows sigmoidally with time. This may indeed be expected as was shown in aforementioned paper (Dobrzyński et al. 2016) in which the sigmoidal dependence on dose resulted from overlapping number of linear dependencies.

It is not easy to calculate how the number of repair enzymes depends on time. However, one can assume that the growth should also be described by a sigmoidal function, so the postulated eq. (5.3) must be modified. Furthermore, since the number of mutations necessary for a neoplastic transformation of a cell is between 2-8 (Vogelstein et al., 2013, Renan 1993; Hahn et al. 1999; Hahn and Weinberg 2002), one can use a formula analogous to eq. (5.4), but with powers 2-8. It may be noted that in order to employ their modelling approach to fitting the Cohen's (1995) data on lung cancer vs dose, Fleck et al. (1999) used the power $m = 5$, which seemed optimal.

The number of mutations in a cell, $m$, is critical for a possible neoplastic transformation to occur. One can assume that the probability of this transformation is 1 at, say, 10 mutations, and may depend on the number of mutations in sigmoidal fashion using the Avrami-Mehl equation (Dobrzyński et al. 2016) as:

$$P_c(m) = 1 - e^{-0.0277 m^k} , \qquad (5.5)$$

which for $m = 5$ and $k = 2$ is close to 0.5, and saturates quickly to 1.0 at larger $m$ values. Obviously, that form of the sigmoidal curve with its *ad hoc* assigned parameter value, as proposed here in eq. (5.5), does not follow from any first principles (Renan 1993; Hahn et al. 1999; Hahn and Weinberg 2002). Generalizing eq. (5.4) to the case of $m$ mutations:

$$M_m = M_0 (1 - e^{-B \, P_m \, t})^m , \qquad (5.6)$$

Using eq. (5.5) one should get some estimation of the number of cells with $m$ mutations that transform to cancer cells:

$$N_{canc}(m,t) = M_0(1 - e^{-BP_m t})^m \left(1 - e^{-0.0277 m^k}\right). \tag{5.7}$$

Eq. (5.7) does not take into account any cooperative action within a collection of cells. It relates only to the creation of cancer cells from the mutated ones. As an example, Fig. 4 shows contour plots in the coordinate system *m-t* for $B \cdot P_m$ = 0.01 and for the exponent *k* in the eq. (5.7) equal to 2 and 4.

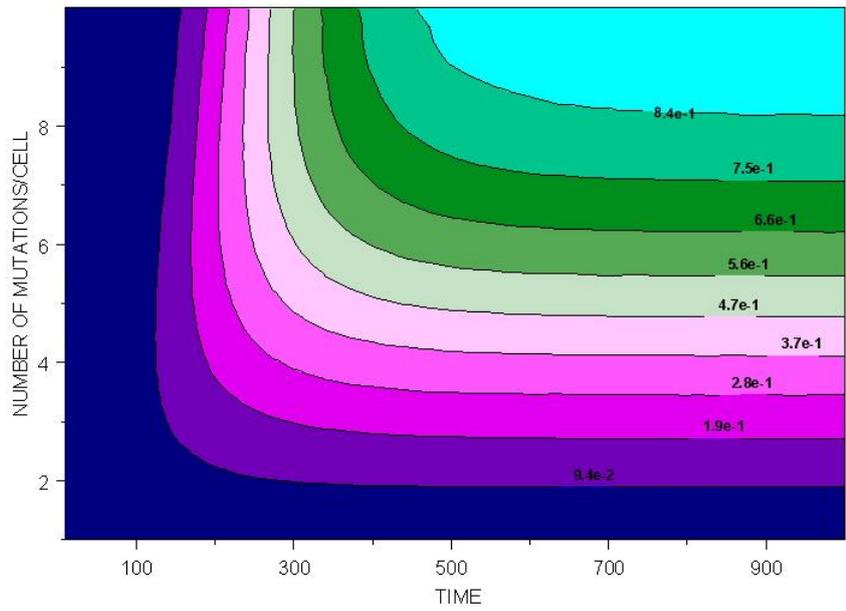

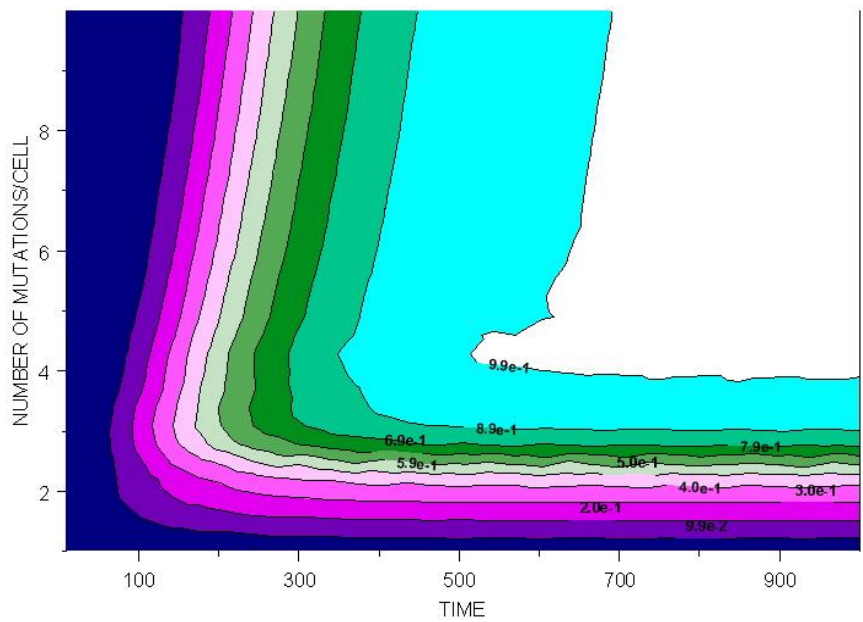

Fig. 4 Relative number of cancer cells vs time, *t*, and the number of mutations per cell, *m*, for various critical exponents $k = 2$ (upper figure) and 4 (lower figure).

The development described by eq. (5.7) must terminate when the number of cancer cells, i.e. the sum of $N_{canc}(m,t)$ over $m$:

$$N_{cancer}(t) = \sum_m N_{canc}(m,t) ,  \qquad (5.8)$$

attains some critical value at which the voluminous tumour growth starts. Let us denote the time at which such a situation happens by $t_{cr}$. Fig. 5 shows $N_{cancer}(t)$ calculated under the assumption that the factor $B$ is constant (independent of $m$) which to our understanding may be the case. Fig. 5 shows that the calculated proliferation rate of cancer cells with time increases with the increasing critical index of cancer growth. This is reasonable as the increase of the critical index means that the rate of transformation to a cancer cell must rise. In all cases the curves $N_{cancer}(t)$ in Fig.5 exhibit a saturation and resemble the sigmoidal Gompertz curves. Of note, this saturated value, after summing up contributions from all values of $m$) can be calculated as:

$$N_{cancmax} = \lim_{t \to \infty} \sum_m N_{canc}(m,t) = \sum_m M_0 \left(1 - e^{-0.0277 m^k}\right) \qquad (5.9)$$

In a special case when $m \geq 4$ and $k \geq 4$, one can write that $N_{cancmax} \approx m \cdot M_0$. The curves in Fig.5 are qualitatively similar to the ones obtained by an analytical approach of Dobrzyński et al. (2016), see their Fig.3. The shape of all of the curves is virtually identical, differing only by a multiplication factor. These curves are, however, quite different from the ones obtained by Fornalski et al. (2011) who used Monte Carlo simulations of the cancer cells' growth. As mentioned earlier, these curves can be perfectly described by the Gompertz curve; Fig. 6 shows the fit of the Gompertz curve $N_{cancer}(t)=8.27844 \cdot \exp[-6.51319 \cdot \exp(-0.010028 \cdot t)]$ to the calculated points for the exemplary case of $k=4$ from Fig. 5. To the best of the authors knowledge, this is the first demonstration of the Gompertz curve (which traditionally describes the time of growth of cancer cells (Laird 1964)) to be obtained from the combination of the probabilities and the basic biophysical properties considered in this paper. (It is particularly noteworthy that the presented calculations, especially Eq. (5.8), do not take into account the processes of cell divisions and deaths that could modify the curves $N_{cancer}(t)$.)

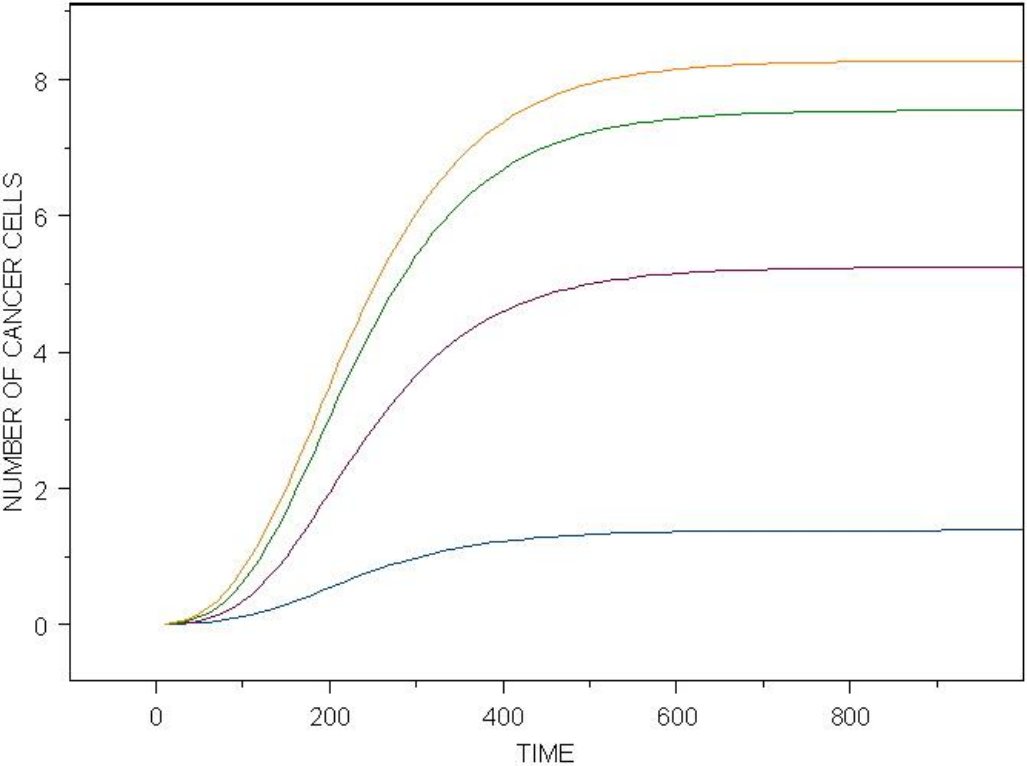

Fig. 5 Number of cancer cells vs time. The curves in ascending order (from brown to blue) correspond to $k$ = 1, 2, 3 and 4, calculated using eq. (5.8) and the assumption of $B \cdot P_m = 0.01$.

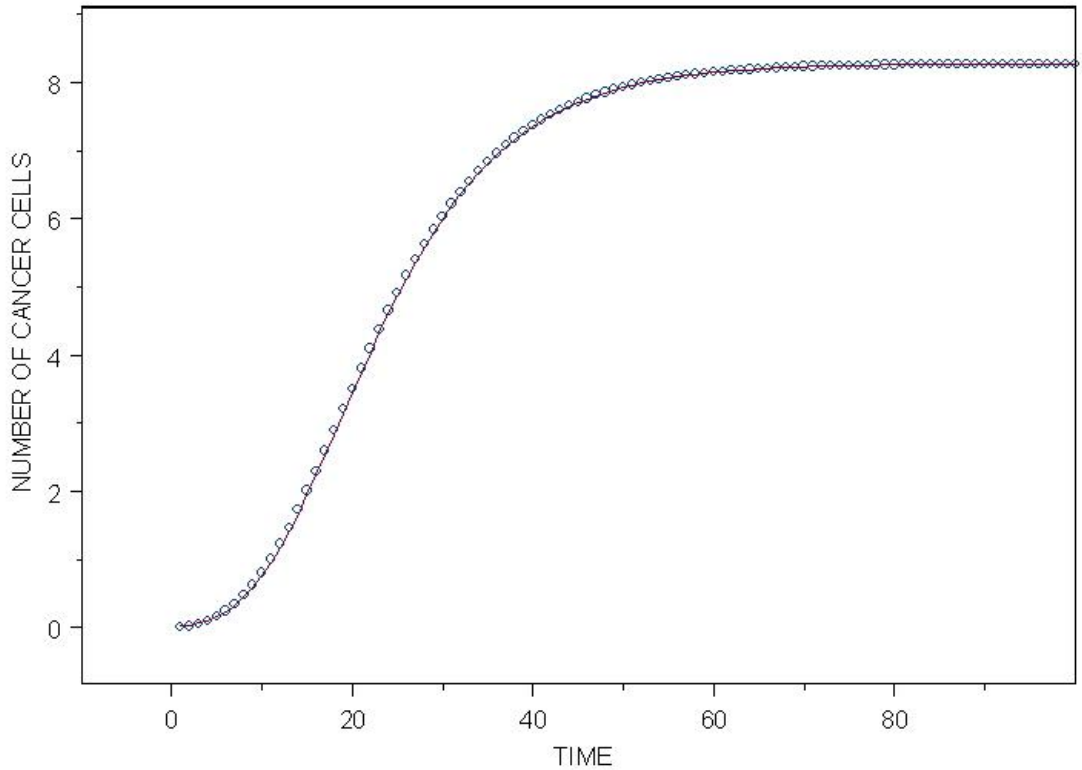

Fig. 6 Calculated number of cancer cells vs time (open circles) from the special case of *k*=4 from Fig. 5, with *B·P$_m$*=0.1. The fitted Gompertz function *N$_{cancer}$(t)* = *8.27844·exp[-6.51319·exp(-0.010028·t)]* (solid line).

As was shown by Dobrzyński et al. (2016), eq. (13), the tumour growth can be also described by the Mehl-Avrami type of equation, which is based on the nucleation and growth theory. It is important to understand that '*it takes a tissue to make a tumour*' (Barcellos-Hoff 2001), and that '*the cancer induction is more a function of the tissue response and not a single cell response*' (Puukila et al. 2017). Thus, consideration of what happens in specific cells rather than in a whole tissue is not sufficient. The tumour growth over time is governed by a critical index *n* showing the spatial type of the growth: linear, 2D or 3D ($n=2$, 3, and 4, respectively). Finally, the achieved tumour volume expressed in terms of the number of cells is characterized by:

$$V = N_{cancmax}\left[1 - e^{-\gamma(t-t_{cr})^n}\right], \qquad (5.10)$$

where $N_{cancmax}$ denotes the number of cancer cells in this volume, so $N_{cancmax}$ corresponds to the maximum given by eq. (5.9), see Fig. 6. Similar reasoning was used by Laird (1964) who originally connected the Gompertz function with tumour growth. The coefficient $\gamma$ must be proportional to the dose rate with the same power index, *n*, so the argument of the exponent in eq. (5.10) is directly connected with the dose as indicated by Dobrzyński et al. (2016).

One can wonder when exactly begins the growth of a 3-dimensional cancer. This may be just a singular (critical) point as is common in phase transitions or in catastrophes. Transition to a self-organized state may also be considered. Whether or not this relates in any way to the self-organized criticality theory (Bak 1996) is not yet clear.

In fact, it can be assumed that in a multistep process of carcinogenesis (Hanahan and Weinberg 2000), each step marks a phase transition. Consequently, the 'cancer energy' landscape of a biological system can be represented by a multitude of energy valleys separated by potential barriers.

The choice of the critical index, *n*, is not trivial; however, it seems reasonable to limit it to 4. Because of the complex nature of a tumour growth this index may not even be an integer. Once again, it is crucial to recognize whether one is considering the acute or chronic radiation exposure.

6. *On cancer growth*

Modelling carcinogenesis is an extremely difficult task because of the multitude and diversity of cancers as well as their many biological and geometrical features that have to be taken into account. As an example, one can consider a hypothetical case of a spherical tumour which obtains its nutrients from the surrounding tissues before the development of its own vascular system. The nutrients enter the tumour by diffusion and their supply decreases with time. A solid tumour itself can contain the inner sphere of dead cells and the outer shell of live cancer cells, both quiescent and actively proliferating (e.g., Aguda and Friedman, 2008, La Porta and Zapperi, 2017). In a slightly more mathematically complicated model than the ones considered heretofore by us, after neglecting the shell of the quiescent cells and assuming a constant rate of the nutrients' consumption, one finds (Foryś, 2005) that the volume of the tumour changes with time according to equation:

$$\dot{V} = \frac{kV}{\gamma}\left[1 - \left(\frac{V}{\vartheta}\right)^{\gamma}\right], \qquad (6.1)$$

where $\gamma=2/3$, $k=2a(c_z-c_a)/3$, $a$ is a scaling constant, $\vartheta = \frac{4\pi}{3}\left[\frac{15}{G}(c_z - c_a)\right]^{3/2}$, where $c_z$ denotes the concentration of nutrients available outside of the tumour, $c_a$ is a constant that relates to apoptotic cell death, and $G$ denotes the rate of the consumption of nutrients. If not for the exponent $\gamma$, the eq. (6.1) would be identical to a logistic equation (if $\gamma = 1$) with limiting value of the volume.

A more advanced approach to spherical tumours is presented by Jiang et al., 2005. Valuable reviews of other analytical models of tumour growth can be found in the already cited book by Aguda and Friedman (2008) and a recent monograph by La Porta and Zapperi (2017). Most of these models describe in mathematical terms the biology of cancer formation. In this respect, equation (6.1) and the ones discussed by us in the previous sections present simplified approaches to this very complicated problem. A multiscale model of avascular tumour growth was also considered in detail in the book by Aguda and Friedman (2008). By using the real data on the colorectal cancer, which has a spheroid shape, these authors showed excellent agreement of their experimental results with those of the Monte Carlo calculations presented by Jiang et al. (2005) in their Fig.5. The time dependence of the elementary volumes of clusters has been assumed to fulfil special requirements related to the capacity of cell division. As a result, the growth of the tumour volume turned out to be fairly well described, while the growth curve could be fitted with the Gompertz function. On the other hand, the solution of eq. (6.1) leads to time dependences with shapes similar to the logistic curve as well as to the one showed in our Fig. 6.

The models presented thus far may be useful in characterizing the time dependence of the carcinogenic process. They rely on the following simplistic reasoning: the DNA in cells is attacked by ionizing radiation (which is our focus) as well as by free radicals (produced during normal aerobic metabolism) which evoke lesions in the DNA structure. If unrepaired, these lesions may be passed on to the next generation cells and give rise to mutations which, when expressed in proto-oncogenes. and tumour suppressor genes may lead to neoplastic transformations of cells.

As indicated earlier, the existence of radio-adaptive responses induced by low-dose irradiations invalidates LNT model employed as a basis for radiation protection regulations (Scott, 2017). This is because radiation doses used to demonstrate the adaptive response (a small dose followed by a large dose) are not additive as required by the LNT model. Moreover, according to this model, potential mutations and neoplastic transformations caused by absorption of low radiation doses add to the number of spontaneously produced mutations and transformations. However, actual data show that exposures at small radiation doses down-regulate rather than increase the amount of such spontaneous effects (Koana et al., 2007; Ogura et al., 2009; Scott 2014, 2017).

It is now commonly accepted that cancer indeed arises from a single cell transformed through a series of genetic mutations, epigenetic events, and environmental determinants that cause and sustain ectopic expression of growth-related genes (see the reviews by Kreso and Dick (2014) and Islam et al. (2015)). The cardinal property of this single cell is its 'stemness', i.e., the capacity for self-renewal and multi-lineage differentiation into subclones of daughter cells which, after further genetic and epigenetic changes, produce heterogenous populations of cancer cells that shape the complex ecosystem of each neoplasm (Allison and Sledge, 2014).

Let's look at this evolving ecosystem of cancer from another angle. It is clear that by passing from single cells to tissues and to organs the organization of the system changes significantly. Since one deals with a complex system (Rubin, 2017), description of the changes should closely follow the rules of phase transitions (e.g., Stanley, 1971) and of self-organization and complexity (e.g., Kauffman, 1993, Heylighen, 2008). As pointed out by Heylighen (2008) 'complex systems consist of many (or at least several) parts that are connected via their interactions. Their components are both *distinct* and *connected*, both autonomous and to some degree mutually dependent.' The other feature of a complex system is that its main units (in this case, cancer cells) are free in the sense that they can multiply or die. Obviously, this description fully reflects cancer development in the environment of normal cells and the extracellular matrix. Notably, however, a rather fundamental question of the nature of cancer cells has not been answered so far. According to Soto and Sonnenschein (2005) a cancer represents a problem of tissue organization in which emergent phenomena (characteristic for complex systems) are of primary importance. Likewise, Mansury et al. (2002) highlight the fact that 'linear adding-up of individual cell behaviour is invalid in the presence of the hypothesized nonlinear interaction among tumour cells, and their environment. Moreover, nonlinearity would render it virtually impossible to predict the long-run dynamics of the system using a purely analytical approach.' Consequently, one must expect the discontinuities in a description of the transitions "from a photon/particle to a cell to a tissue and to a cancer." In fact, even on the level of genes and cells one observes emergent phenomena and the self-organized criticality (e.g., Tsuchiya et al. 2015). Notably, Mansury et al. (2002) who claim that 'malignant tumours behave as complex dynamic self-organizing and adaptive biosystems' also indicate that distinct phase transition properties can be found in the number of cancer cell clusters and their temporal behaviour vs. the intrinsic capability of a single cancer cell to migrate. In this context, another important question is whether the once formed cancer cell stays as such until its death (Sonnenshein and Soto, 2016). Apparently, none of the earlier discussed analytical approaches addresses this question. One can also point out the remark by Prehn (1994) that a cancer may not be caused solely by mutations in the DNA and a cancer cell may not stay as such forever (i.e., can reverse to a normal cell state). Sotto and Sonnenschein (2005) suggest therefore that "it may be more correct to say that cancers beget mutations than it is to say that mutations beget cancers."

In their seminal papers, Hanahan and Weinberg (2000; 2011) indicate that during carcinogenesis neoplastically transformed cells acquire critical features called 'the hallmarks of cancer'. These include: growth factors self-sufficiency, insensitivity to anti-growth signalling, evasion of programmed cell death (apoptosis), limitless replicative potential, sustained angiogenesis, ability to invade and metastasize, genome instability and enhanced mutation rate, reprogramming of the energy metabolism, and evasion of immune destruction. Additionally, Frederica Cavallo and co-workers (2011) proposed two 'immune hallmarks,' i.e., the ability of cancer cells to thrive in a chronically inflamed environment and to suppress immune reactivity.

At the beginning of the 21$^{st}$ century Schreiber and his colleagues described a process called 'cancer immunoediting' whereby the immune system, the most potent guardian against neoplasia, prevents cancer development at the early stages of carcinogenesis, but also shapes ("edits") immunogenicity of neoplastic cells and contributes to cancer development (Shankaran et al., 2001; Dunn et al., 2002; Shreiber et al., 2011).

The cancer immunoediting process can be divided into three consecutive phases: 1. *elimination*, during which incipient cancer cells are recognized by the alarmed innate immune system which triggers adaptive immune responses that specifically detect and destroy neoplasticcells; 2. *equilibrium*, when humoral and cellularimmune mechanisms (e.g., interferon-$\gamma$, interleukin-12, granulocytes, macrophages, T and B lymphocytes) hold persisting cancer cells in check (*cancer dormancy*), but also shape the immune status of these cells and their environment; and 3. *elimination*, during which the extant and "immunoedited" (i.e., resistant to immune attack) cancer cells proliferate in the immunosuppressive environment facilitating cancer progression toward a full-blown, clinically detectable disease (reviewed in Janiak et al. 2017).

From the perspective of this paper the most important is the third phase of cancer development. Apparently, during all phases of cancer immunoedition there is a competition between stimulation and inhibition of cell proliferation, and between dynamic disorder and order. Hence, the use of a deterministic approach, as we did in earlier sections and in eq. (6.1), cannot satisfy the needs: our system is non-linear, and should not be described by linear equations as stated explicitly by Mansury et al. (2002). Moreover, there is little hope that a reductionists point of view will help us to comprehend how a cell is functioning until we fully understand the variety of molecular interactions in a normal cell and how cells function within normal tissue, an organ, and a cancer. As indicated by Saetzler et al. (2011) 'he upward causation assumption completely neglects the contribution of the environment and of the emergent structure itself (by downward causation).' The popular somatic mutation theory (SMT) of cancer turns out to be insufficient to explain the variety of cancer behaviour. A tissue organization field theory was proposed to better accomodate and explain the emerging experimental evidence related to cancer development (Sonnenschein and Soto, 2016). With regard to phase transition in a complex system, the natural question is what happens close to the point at which phase transition takes place? To what extent can such a transition be treated as an emergent phenomenon? Is the loss of control over tumour growth a sort of catastrophic event, such as an avalanche (Bak, 1996) treated as an indicator of self-organization at the phase-transition, or is it just a phase transition of the first or second order, similar to the re-entrant and other transitions of frustrated systems from disordered to ordered states (Binder and Kob, 2011). In the case of self-organization an essential difference between what is going on before and after the transition relates to the fact that before the transition every cell more or less individually interacts with its closest tissue constituents. In contrast, after the transition, all cells work together: what happens in one place of the tumour has a direct influence on what happens at any other place. How the situation in one place will change the situation in another place of the tumour is hard to predict (possibly abscopal effects may be involved). At the beginning, all incipient tumour cells are fed by diffusion from the surrounding tissues and can proliferate. With time, however, the inner core of a tumour is formed. Probably, eq. (6.1) may roughly describe the evolution of a tumour volume, mainly of its outer shell. From the organization point of view, the larger the tumour, the more external, unbounded cancer cells can be accommodated on its surface, and the tumour would exponentially grow up to infinity. In any case, within the scope of the theory of complexity one has to admit that the process of tumour growth cannot be reduced to individual interactions between cells as was possible during the equilibrium phase of the immunoediting process. Once the tumour is formed after passing the critical point, the tumour cells lose their individuality and become totally subordinated to the properties of this new entity. Of course, any stress, such as exposure to ionizing radiation, may change this self-organized behaviour. Because of such complexities, one has to accept that the description of an organism cannot be reduced to interactions between its principal entities and that the organisms are subject to rules of organization and its variations, as discussed by Mossio et al. (2016). From a purely physical point of view an organism is an open system capable of exchanging energy and matter with its environment. These important problems are, however, beyond the scope of the present paper.

Phase transition as discussed here may be also well understood based on the so-called percolation-type of phase transition. In this case, the main assumption is that a single cancer cell is not a cancer itself, and only a group of these cells may constitute a tumour. So, at first, individual cancer cells may occasionally form contacts (or links) between each other and the functional links may lead to the formation of clusters. Within the scope of the continuous time branching processes theory one can calculate a cumulative distribution of the cancer colony sizes vs. the number of cells in these colonies (see Fig. 3.3 in La Porta and Zappero, 2017). This distribution depends on the time of observation, qualitatively is not very different from the logistic curve, and pretty well describes the observations.

In a further development of such an intertwined network, all the clusters may fill up the space in such a way that nutrients provided to one of them can be transferred, to any other – the percolation transition is achieved. Likewise, a disturbance occurring in one place can be propagated to any other place. In terms of the second order phase transitions one should talk about spatial and temporal fluctuations that grow, on average, below the transition point and, upon passing this point, freeze to a single ordered phase.

As mentioned in the previous section, a single cancer cell created during the neoplastic transformation of a mutated cell is not yet a cancer. Eqs. (5.10) and (6.1) described the tumour growth with time. To make a group of cancer cells a tumour, a certain number of them, $h$:

$$h = \frac{N_{canc}}{N_{canc}+N_{non\_canc}}, \quad (6.2)$$

where $N_{non\_canc}$ is a sum of damaged and mutated cells ($N_{non\_canc} = N_{lesion} + N_{mut}$), must interact with each other and start to proliferate in a coherent way after passing a certain critical value, $h_c$.

With time, the value of $h$ changes, but as long as cancer cells or their clusters are disconnected the tumour has not yet emerged as a separate entity. As already indicated, such an emergence can be identified with a phase transition similar to the re-entrant phase transitions known from the physics of magnetics or the percolation theory. In both cases, the final formation of a given object (e.g., a tumour) appears when $h$ exceeds some critical value (e.g., the number of cancer cells, $h_c$), as mentioned earlier. Within the scope of the percolation theory, which refrains from purely physical or biological parameters, one needs to define the parameter which should control the occupancy of pixels (voxels) into which a given space is subdivided. This parameter should reflect not the relative number of cancer cells as in (6.2) but rather the probability of the creation of a cancer cell at a given pixel (voxel). Let this probability be denoted by *p*. With time cancer cells aggregate and form clusters that combine with each other and finally a critical state is attained: the 'infinite cluster' (a tumour) is formed in which any information sent from one location in the object can reach all other locations with a consequence to the whole object – the tumour starts to behave as an entity whose behaviour cannot be derived from individual properties of cells and their interactions. This is similar to the sand-pile experiment (Bak,1996): although the grains of sand drop to the sand's cone from the top onto a single point, at a certain height and radius of the cone avalanches appear in an unpredictable manner (the so-called self-organized criticality). In the case of percolation the situation may be visualized by imagining a number of pixels or voxels into which one drops small grains or small spheres. Next let us connect randomly any two such small spheres. If we repeat this procedure, the number of the connected spheres will increase, and the number of locally connected spheres (clusters) will increase. At a certain moment the size of those clusters will start to suddenly rise. This illustrates a case of self-organization.

In a typical simulation of a percolation phenomenon (Binder and Kob, 2011) one has to choose the size of the object divided into pixels, and by the 'infinite cluster' one understands the cluster extending from one edge of the object to another one. Ideally, the object should have infinite dimensions, so that the meaning of 'infinite cluster' is literal. In our case the situation is somewhat different. An organ is close to be infinite with respect to the size of the cells. The size of tumours in it may not be as large as the organ for, e.g., the lack of nutrients needed the cancer to grow (Laird 1967).

Nevertheless, one can still treat the maximum size of the tumour as roughly equivalent to an infinite lattice of cells and the percolation theory (Stauffer, 1979) with its purely geometrical statistical ingredients can be useful for the description of at least a region close to the phase transition point, $p_c$. In the case of a site percolation on the square lattice $p_c \approx 0.593$. The 'infinite cluster' may exist only above $p_c$. Inside the organ, which represents a truly infinite lattice, one can imagine formation of more than one 'infinite cluster', i.e., of more than one tumour.

According to this theory, the probability that a given cell belongs to the collection of tumour cells grows at $p > p_c$ in a critical way, and the percolation probability, $P_{max}$, i.e. the fraction of the occupied sites belonging to the 'infinite cluster' is ruled by the critical index $\beta$ which may in general be of the fractal type:

$$P_{max} \sim (p - p_c)^\beta \ . \qquad (6.3)$$

This behaviour is displayed in Fig.6. It is neither sigmoidal type nor logistic type, however one should remember that the relation (6.3) must describe the behaviour mainly in the critical region, i.e. relatively close to $p_c$. Alas, the width of the critical region is difficult to predict.

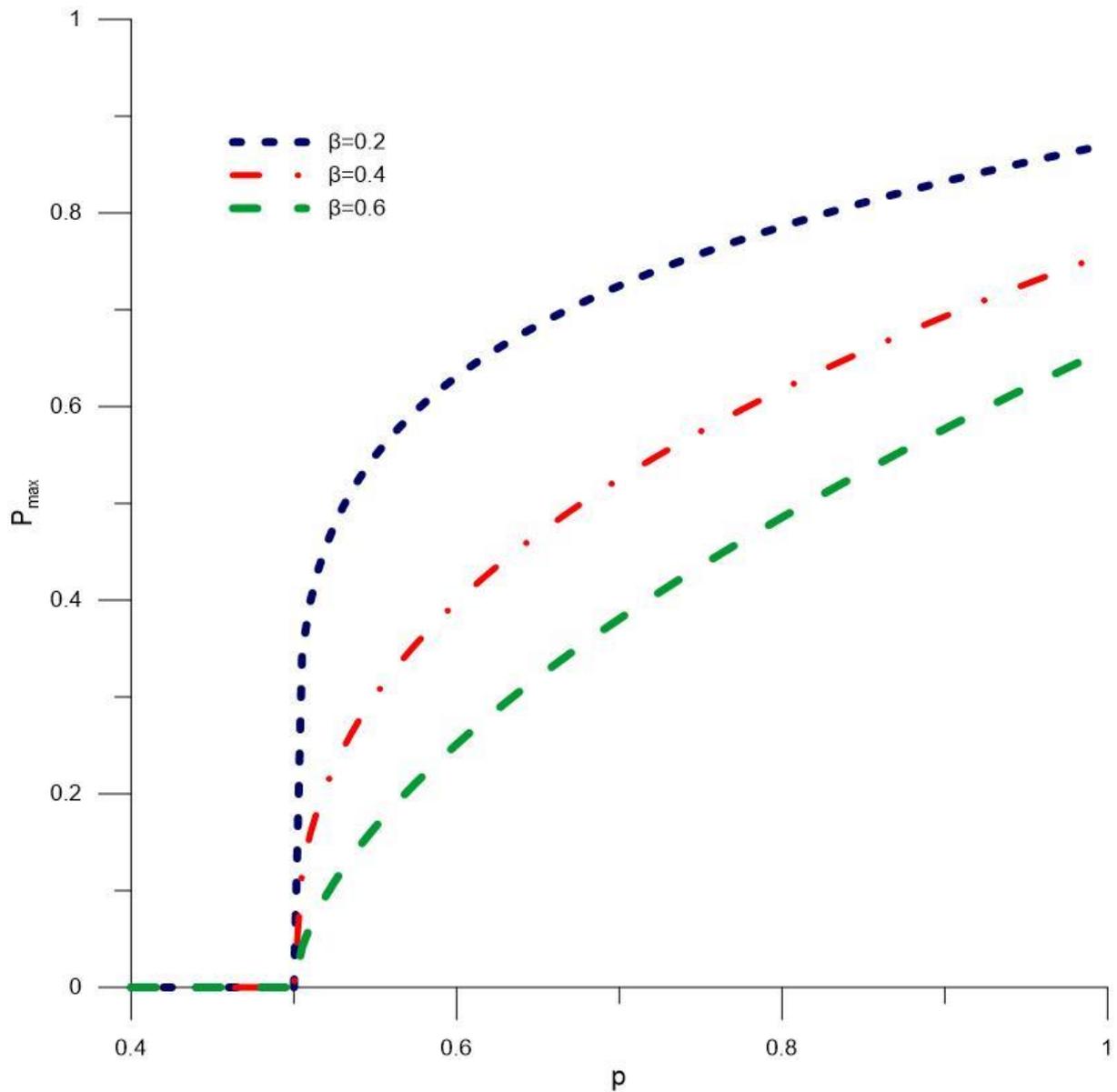

Fig.7 Transition to percolation after passing the $p_c$ point assumed to be $p_c = 0.5$ at different critical indices $\beta$.

A very illustrative example of this kind of behaviour can be found in many disordered magnetic systems, such as a diluted ferromagnet which below $p_c$ becomes a paramagnet, and becomes ferromagnetic above $p_c$. In addition, over a certain concentration range of magnetic species (Co, Fe, …) the spin-glass phase can be formed. In such a system the control parameter is temperature – with increasing temperature the value of the percolation probability $p_c$ increases as well. As an example, the onset of ferromagnetism in a percolating 3D network of the random fcc (face centered cubic) alloys can be satisfactorily described within the framework of the percolation theory, where $p_c$ ranges from 0.16 to 0.20 (Childress and Chien, 1991). Often the phase diagram of such diluted ferromagnets can be described within the framework of the so-called Ising model. In the case of cancer one encounters a more complex situation because the number of states representing cells with different degrees of lesions and mutations is much higher than the number of up and down spins in Ising model. This makes statistical description of the state below $p_c$ more difficult.

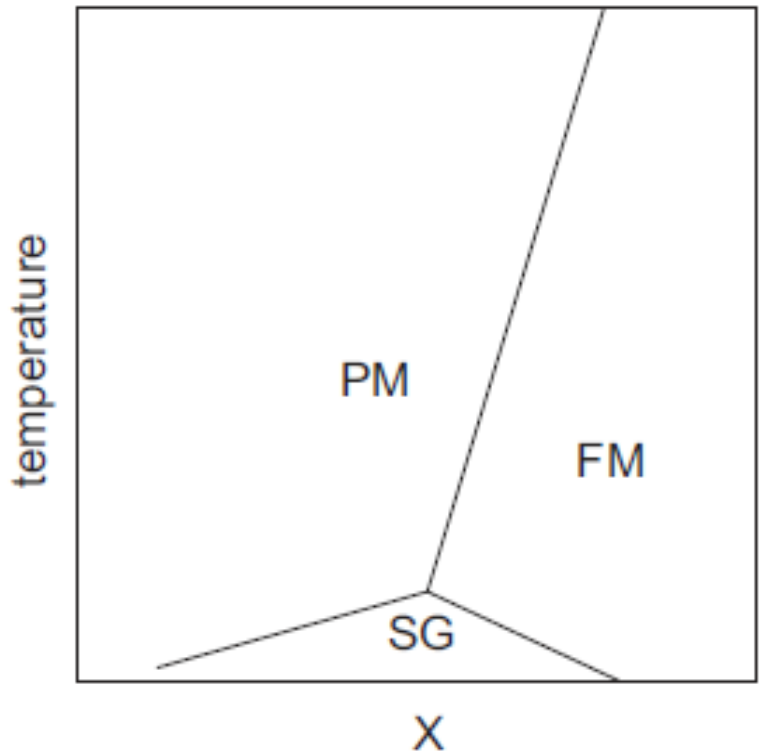

Fig.8 Schematic phase diagram of a diluted ferromagnet (e.g. Au-Fe, Co-Cu etc.) showing phase transitions between paramagnetic (PM), ferromagnetic (FM) and spin-glass (SG) phases depending on temperature and concentration X of the magnetic sample.

It should be mentioned that the sharp transition shown in Fig.7 may be smeared if one finds correlations between the cells below $p_c$. If such correlation exists between the states of a two cells separated from each other by a distance $R$, one can find a function describing such a correlation. In a typical magnetic system it would be described by a function decreasing exponentially with the distance. This would lead to a substantial change in Fig. 7: the whole curve becoming sigmoidal-like, such as the one in Fig. 3.9 of Binder and Kob, 2011.

Just below the transition point the average distance between the cells within a cluster (correlation length), ξ, behaves as:

$$\xi \sim (p_c - p)^{-\nu}, \qquad (6.4)$$

where ν denotes another critical exponent. The values of both exponents depend on grid and slightly differ depending on case[2]. For example, for a 3-dimensional net β = 0.418 ± 0.001, and ν = 0.875 ± 0.008. The mean cluster size also exhibits non-analytical behaviour:

$$S \sim (p_c - p)^{-\gamma}, \qquad (6.5)$$

with γ = 1.793 ± 0.003. In the case of two-dimensional growth the critical exponents change to 5/36, 4/3 and 43/18 for *β, ν,* and *γ,* respectively. Let us note that the critical exponents depend on the dimensionality of the problem and not on the microscopic properties of the system. The whole situation resembles the behaviour of magnetic systems: our *p* plays a role similar to temperature, percolation probability $P_{max}$ – magnetization, ξ - correlation length, and *S* – magnetic susceptibility. Below the transition point one can also make intuitive use of the so-called mean-field approach. Namely, let the number of cancer cells be $N_{canc}$, whereas $N_{canc,0}$ is the number of cancer cells whose growth seem to be inhibited by immunological forces and represent dormant cells. The development of cancer cells may depend on their number that form a special field due to interaction of these cells with their environment (in fact, it is the question of competing forces between natural expansion of cancer cells and the immune restrain of such a development, see Janiak et al, 2017). This field describes something analogous to a promotion of the development of cancer with a multiplication factor, say, $\chi$. Then one can write a simple equation:

$$N_{canc} = N_{canc,0} + \chi N_{canc} . \qquad (6.6)$$

---

[2] https://en.wikipedia.org/wiki/Percolation_critical_exponents#Exponents_for_standard_percolation

Thus;

$$N_{canc} = \frac{N_{canc,0}}{1-\chi}, \qquad (6.7)$$

which shows how the number of cancer cells can be strongly enhanced by the interaction of these cancer cells with their environment. In an extreme case the number of cancer cells can reach the total number of cells if $\chi = 1$, i.e., when immunological protection against cancer development breaks down. Eg. (6.7) is a typical result obtained within the framework of the mean-field theory of phase transitions. The problem to be solved is the description of the form of $\chi$. It is not a single number but a function depending on the processes of immunological protection against cancer. In fact, all the aforementioned steps in cell behaviour represent some phase transitions although an appropriate mathematical description of these transition is rather difficult. Consequently, one needs to define the so-called order parameter which changes upon transition. This parameter must have defined dimensionality, while one should also define whether the main interaction mechanism with environment is short- or long-ranged. Besides, it seems that one deals with three space-time scales: the one of signal transduction, the short-range adaptive response and bystander effect, and the long-ranged abscopal effect.

Presently, the above described critical phenomena are difficult to observe at present as the minimum size of the detectable tumour is of the order of a few milimeters, when the tumour is already formed. In spite of this obstacle, one should understand that the general properties of phase transitions, including self-organization and/or self-organized criticality have to be included in a rigorous description of the tumour development.

7. *Conclusions*

There are several problems covered in this paper. Its intention was to separate the evolution of individual cells from that taking place in a tissue. With this aim in mind, the approach of Fleck et al. (1999) was initially examined. Although the ideas behind their model are quite different from the ones presented previously by us (Dobrzyński et al. 2016), the final result for the number of cancer cells vs time has turned out to be qualitatively similar. Notably, in contrast to Fleck et al. (1999) who used dose rate, Dobrzyński et al. (2016) employed dose in their calculations. Moreover, a more fundamental difference in the way a cell repairs its DNA lesions is presented in the presented paper. In the model of Fleck et al. (1999) the dependence of the time of repair of a single DNA lesion on the dose rate plays an essential role. Apart from the fact that a postulated modification of the repair time with increasing dose rate may be described by a different function, the model of Fleck et al. (1999) explains Cohen's (1995) data on lung cancer mortality vs radon-specific activity. However, it seems that one cannot go too far with a function fitted to data, as the mortality calculated in this way soon exceeds the hormetic minimum. Such an increase is not confirmed by the numerous other data collected for much higher radon concentrations (Dobrzyński et al. (2018), (Henriksen, 2015).

Our attention has also been paid to the dependence of adaptive response on the time and the dose, associated with acute or protracted radiation exposures. We demonstrate that in the case of a protracted exposure, the organism attains a certain saturation in its ability to repair lesions (Fig. 5). This observation permits us to treat absorbed dose as a priming dose which allows to better tolerate higher and more challenging doses.

Increase in the number of cancer cells in an organism depends on the timing and the dynamics of the critical number of mutations in a cell needed for its transformation into a cancer cell. As demonstrated by us, this may be described by a pretty complicated function (5.7). However, the increase in the total number of tumour cells (Fig.5) when graphed, resembles a sigmoidal shape of the Gompertz function. As previously indicated by Dobrzyński et al. (2016), the 3-dimensional growth of the tumour volume may generally be described by a sigmoidal curve given by the nucleation and growth according to the Avrami-Mehl theory as discussed in Dobrzyński et al. (2016).

Finally, we have discussed at length the problem of tumour growth. Description of this process is inherently difficult. Firstly, it essentially deals with the phase transition phenomenon, and the type of this transition is not clear. Essentially, there is a transition from a disordered phase (represented by individual cancer cells and their clusters) to an ordered one (the developed tumour) which exhibits properties not directly connected to individual processes occurring in its basic units (cells). If this is the case, one is dealing with an emergent phenomenon, the self-organization and the possible self-organized criticality. However, even if the phase transition can be viewed as a continuous one (of second order) it is difficult to specify its most important parameters: the order parameter and its dimension, as well as the interaction range between cells, and the control parameter (usually temperature in a magnetic system or in, e.g., liquid-gas transition). Depending on these parameters one could specify the values of critical exponents characterizing the transition. An important property of such critical indices is their universality – they are not directly related to microscopic interactions between the basic units of the system. However, when a mathematical description is given, its experimental verification can be difficult. Today, a tumour cannot be detected if its diameter is around 2-3 mm. This means that it is already far beyond the phase transition region where critical properties can be observed. To illustrate the problem, phase transitions of the percolation type were considered. Additionally, a simple application of the mean field theory shows that non-analytical behaviour is to be expected if a phase transition takes place. We plan to address these problems in a future publication.

*Acknowledgements*

*The authors are greatly indebted to Dr. Bobby Scott, prof. Michał Waligórski and prof. Mark Kon for their numerous remarks and improvements in this paper, and also to Dr. Yehoshua Socol and Dr. Nicholas Keeley for reading and making useful comments on the manuscript.*

*Appendix A*

In the case of a single photon of gamma or X-ray radiation, one must consider cross sections for the photoelectric effect ($\sigma_{ph}$), for Compton scattering ($\sigma_C$), for electron-positron pair production in the nuclear field ($\sigma_{pair}$), and for similar pair production in the electron field, so-called triplet production ($\sigma_{triplet}$). Thus the total cross section for the interaction of gamma- and/or X-rays with matter can be described by:

$$\sigma(R, E) = \sigma_{ph} + Z\sigma_C + \sigma_{pair} + Z\sigma_{triplet}, \qquad (A0)$$

where *Z* is the atomic number of the absorbing material (note that Compton and triplet effects cross sections are calculated for single electrons). In fact, Eq. (A0) describes the ionization processes that can happen in a cell. The equation written in extended form is rather complicated and will not be presented here – all necessary terms on the r.h.s. of Eq. (A0) can be found in the literature (Hubbell et al. 1980).

*Appendix B*

The basis for the quasi-linear dependence of $P_{hit}$ (Fornalski et al. 2011):

$$P_{hit} = 1 - \exp(-cD) \qquad (B0)$$

stems from a simple observation: let us imagine that single cell is composed of *N* pixels which can be hit by radiation. Some number of them, say *n*, are important from radiobiological point of view and represent e.g. cellular DNAs. Thus, the probability of DNA damage made in a pixel by a single particle hit is *n/N*. In case of two particles impinging on the considered cell this probability changes to:

$$P_{2\,particles} = 2\left(\frac{n}{N} \cdot \frac{N-n}{N}\right) + \left(\frac{n}{N}\right)^2 \qquad (B1)$$

because one shall consider three scenarios: i) first particle hit DNA and the second not, ii) analogical to the previous one, but the opposite, iii) two particle hit DNA. For many (*k*) particles, where some of them hit DNA, one can use the sum of binomial distribution functions as:

$$P_{k\,particles} = \sum_{r=1}^{k} \frac{k!}{r!\,(k-r)!} \left(\frac{n}{N}\right)^r \left(1-\frac{n}{N}\right)^{k-r} \tag{B2}$$

If not the lack of the term with *r* = 0 (none of particles hit DNA), eq. (B2) would be nothing else than binomial of *[n/N + (1-n/N)]$^k$* which is obviously equal to 1. The missing term is

$$P_{0\,particles} = \binom{k}{0} \left(\frac{n}{N}\right)^0 \left(1-\frac{n}{N}\right)^k = \left(1-\frac{n}{N}\right)^k \tag{B3}$$

which should be added to the probability *P$_{k\,particles}$* for *r* ∈ [1,*k*], as in eq. (A2). Thus

$$1 = P_{k\,particles} + \left(1-\frac{n}{N}\right)^k \tag{B4}$$

and

$$P_{k\,particles} = 1 - \left(1-\frac{n}{N}\right)^k \tag{B5}$$

In the case of *n<<N* (which is always correct in our case) the second term on r.h.s of (B5) presents first order expansion of *exp(-kn/N)* (Maclaurin series equation) and finally one finds that

$$P_{k\,particles} = 1 - e^{-\frac{n}{N}k} \equiv P_{hit} \tag{B6}$$

which is the same as *P$_{hit}$* from the eqs (2.2) and (B0), where *c=n/N* represents the probability of DNA hit and *k* (number of particles) corresponds to the dose (dose per numerical step).

The presented approach is analogous to the Target Theory (Lea 1955) applied originally to the survival of a group of irradiated cells.

*Appendix C*

For $t_0 \leq t \leq t_1$ the time-dependent integral term can be written as

$$I(t) = \int_{t_0}^{t} (t-h)^m e^{-\alpha_3(t-h)} dh ,\qquad (C1)$$

and for $t \geq t_1$:

$$I(t) = \int_{t-t_1}^{t-t_0} (t-t_0-h)^m e^{-\alpha_3(t-t_0-h)} dh ,\qquad (C2)$$

which for *m*=1 or *m*=2 is easy to calculate:

for *m* = 1:

$$I(t_0 \leq t \leq t_1) = \frac{1}{\alpha_3^2} - \left(\frac{1}{\alpha_3^2} + \frac{t-t_0}{\alpha_3}\right) e^{-\alpha_3(t-t_0)} ,\qquad (C3)$$

$$I(t \geq t_1) = \left[\left(\frac{1}{\alpha_3^2} + \frac{t-t_1}{\alpha_3}\right) e^{-\alpha_3(t-t_1)}\right] - \left[\left(\frac{1}{\alpha_3^2} + \frac{t-t_0}{\alpha_3}\right) e^{-\alpha_3(t-t_0)}\right] ,\qquad (C4)$$

and for *m* = 2:

$$I(t_0 \leq t \leq t_1) = \frac{2}{\alpha_3^3} - \left[\frac{(t-t_1)^2}{\alpha_3} + \frac{2(t-t_1)}{\alpha_3^2} + \frac{2}{\alpha_3^3}\right] e^{-\alpha_3(t-t_1)} ,\qquad (C5)$$

$$I(t \geq t_1) = \left[\frac{(t-t_1)^2}{\alpha_3} + \frac{2(t-t_1)}{\alpha_3^2} + \frac{2}{\alpha_3^3}\right] e^{-\alpha_3(t-t_1)} - \left[\frac{(t-t_0)^2}{\alpha_3} + \frac{2(t-t_0)}{\alpha_3^2} + \frac{2}{\alpha_3^3}\right] e^{-\alpha_3(t-t_0)}. \quad (C6)$$